\documentclass[a4paper]{article}
\pdfoutput=1
\usepackage{INTERSPEECH2021}

\usepackage[utf8]{inputenc}
\usepackage{bbm}
\usepackage{multirow}
\usepackage{makecell}
\usepackage{cite}
\usepackage{amssymb,subfigure}
\usepackage{placeins}
\usepackage{babel,blindtext}
\usepackage{caption}
\usepackage{tikz}
\usepackage{adjustbox}
\usepackage{verbatim}
\usepackage{hyperref}
\usepackage{xurl}
\usepackage{cleveref}
 
\usepackage[T1]{fontenc}
\usepackage{adjustbox}
\usepackage{mathtools}

\title{Reinforce-Aligner: Reinforcement Alignment Search for Robust End-to-End Text-to-Speech}
\name{Hyunseung Chung$^1$, Sang-Hoon Lee$^2$, Seong-Whan Lee$^{1,2}$ \thanks{This work was supported by Institute of Information \& communications Technology Planning \& Evaluation (IITP) grant funded by the Korea government (MSIT) (No. 2019-0-00079, Department of Artificial Intelligence, Korea University), and the Magellan Division of Netmarble Corporation.}}
\address{
  $^1$Department of Artificial Intelligence, Korea University, Seoul, Korea\\
  $^2$Department of Brain and Cognitive Engineering, Korea University, Seoul, Korea}
\email{\{hs\_chung, sh\_lee, sw.lee\}@korea.ac.kr}

\begin{document}

\maketitle
\begin{abstract}
  Text-to-speech (TTS) synthesis is the process of producing synthesized speech from text or phoneme input. Traditional TTS models contain multiple processing steps and require external aligners, which provide attention alignments of phoneme-to-frame sequences. As the complexity increases and efficiency decreases with every additional step, there is expanding demand in modern synthesis pipelines for end-to-end TTS with efficient internal aligners. In this work, we propose an end-to-end text-to-waveform network with a novel reinforcement learning based duration search method. Our proposed generator is feed-forward and the aligner trains the agent to make optimal duration predictions by receiving active feedback from actions taken to maximize cumulative reward. We demonstrate accurate alignments of phoneme-to-frame sequence generated from trained agents enhance fidelity and naturalness of synthesized audio. Experimental results also show the superiority of our proposed model compared to other state-of-the-art TTS models with internal and external aligners. 
  
\end{abstract}
\noindent\textbf{Index Terms}: text to speech, reinforcement learning 

\section{Introduction}
Rapid progress in text-to-speech (TTS) was initiated by autoregressive models. These models replaced traditional methods \cite{griffin1984signal, yoshimura1999simultaneous, kawahara1999restructuring, suk2011subject, lee2018high} and are typically sequence-to-sequence in an encoder-decoder framework with attention mechanism \cite{arik2017deep, gibiansky2017deep, wang2017tacotron, shen2018natural, li2019neural, li2020robutrans, weiss2020wave}. The purpose of the encoder is to extract hidden representation feature vectors from phoneme sequence, and the decoder generates mel-spectrograms from produced vectors. Despite the advantages, end-to-end attention within autoregressive models have limitations such as slow inference speed, word skipping, and reading \cite{ren2019fastspeech, li2020robutrans}. As a means of remedy to this problem, non-autoregressive models are proposed for parallel generation of mel-spectrograms from text or phoneme \cite{ren2019fastspeech, lim2020jdi, vainer2020speedyspeech, lancucki2020fastpitch, lee2020multi, elias2020parallel, luo2021lightspeech}. Although the new architecture alleviates some of the drawbacks from autoregressive models, the duration aligner of non-autoregressive models still require guidance from external aligners. The most critical issue of having an external aligner is an increase in complexity of the training process. Properly aligned text and speech attention maps are required from an autoregressive model before training. This delays the training process and the non-autoregressive model becomes reliant on the quality of alignments generated by the external aligner. Therefore, recent synthesis pipelines are designed with the objective of robust end-to-end internal aligners \cite{kim2020glow, donahue2020end, miao2020efficienttts, shen2020non, liu2021vara}.

Most recent works on TTS related to our work are EATS \cite{donahue2020end} and HiFi-GAN \cite{kong2020hifi}. EATS is an end-to-end text to waveform network with an internal aligner that approximates phoneme to mel-spectrogram sequence alignments with Gaussian kernels. Although the model proposes robust text-to-wave synthesis, the alignments are not additionally trained to ensure improved duration alignment. Hifi-GAN generates high quality audio waveforms through multi-scale and multi-period discriminators, but the model is limited to considering only mel-spectrograms as input. Our proposed text-to-waveform network represents the best of both worlds and improvement on the drawbacks.

In this paper we propose an end-to-end text-to-wave network with reinforce-aligner, which is a reinforcement learning based alignment search method for robust speech synthesis. Our agent interacts with the environment over a sequence of steps to select the best action given the current state. Then, the environment applies an update to the action and returns the reward of the action for the agent to take into account for the next step. This training process is repeated until convergence, and enables the network to internally learn its own alignment. Our experimental results show the positive impact of the reinforce-aligner on the duration alignment and quality of generated audio waveforms. The synthesized audio samples are provided on our online demo webpage.\footnote{\url{https://prml-lab-speech-team.github.io/demo/Reinforce-Aligner}}                   

\begin{figure}[!t]
\begin{center}
\includegraphics[width=1\columnwidth]{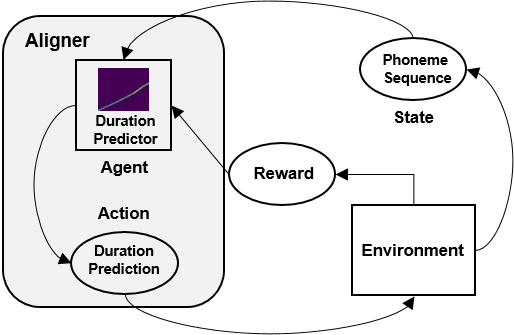}
\end{center}
   \caption{Overall architecture of reinforce-aligner. The agent (duration predictor) interacts with the environment (text-to-waveform network) to receive current state (phoneme sequence) and reward feedback. 
}
\label{fig:1}

\end{figure}

\begin{figure*}[!t]
    \centering
    \subfigure[Generator training process]{
    \centering
    \includegraphics[width=1.50\columnwidth]{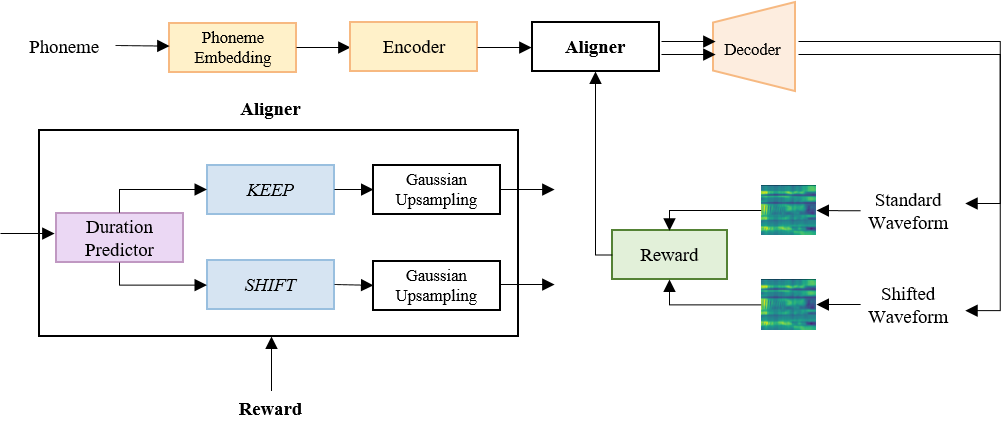}
    \label{generator}
    }
    \subfigure[Reward setups]{
    \includegraphics[width=0.55\columnwidth]{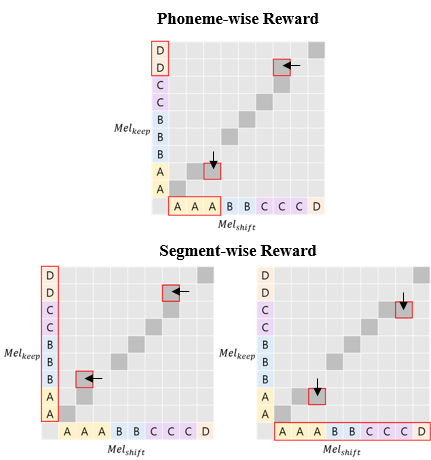}
    \label{rewards}
    }
    \caption{Figure (a) is the overall architecture of the generator in training process. The aligner is shown in detail to describe the actions available for the agent. Figure (b) is the phoneme-wise and segment-wise rewards. Each arrow represents a shift that deviates towards the lower mel-spectrogram loss value.}
    \label{fig:2}

\end{figure*}

\section{Model Architecture}
Our proposed fully convolutional generator utilizes text or phoneme as input and generates a raw audio waveform as output. The encoder contains a modified multi-receptive field fusion module (MRF) \cite{kong2020hifi}. The original implementation of MRF uses mel-spectrograms as input and upsamples the mel-spectrogram with transposed convolution to a raw waveform. Our implementation utilizes phoneme embeddings as input and outputs hidden representations without upsampling. The MRF based phoneme encoder contains multiple residual blocks with multiple kernel sizes and dilation rates, which is an essential component of our network because various receptive fields are able to extract distinct contextual features from phoneme embeddings. Then, the encoder output progresses through the reinforce-aligner to produce output frames. The output frames are randomly segmented by $\gamma$. Finally, the decoder upsamples $\gamma$ by 256 to produce raw audio waveforms. 

 Our model contains two discriminators for different objectives during training. The first discriminator is the Multi-Scale Discriminator proposed in \cite{kumar2019melgan}. This discriminator effectively learns different frequency components of audio through variation in scales. Each of the three sub-discriminators contain convolution on different scales: raw audio, downsampled by factor of 2, and downsampled by factor of 4. The second discriminator is Multi-Period Discriminator \cite{kong2020hifi}, which captures the distinct features of audio with convolutions in periodic variations. Each of the different periodic values consider different periodic segments of the input audio.  

\begin{figure*}[!t]
\begin{center}
\includegraphics[width=180mm]{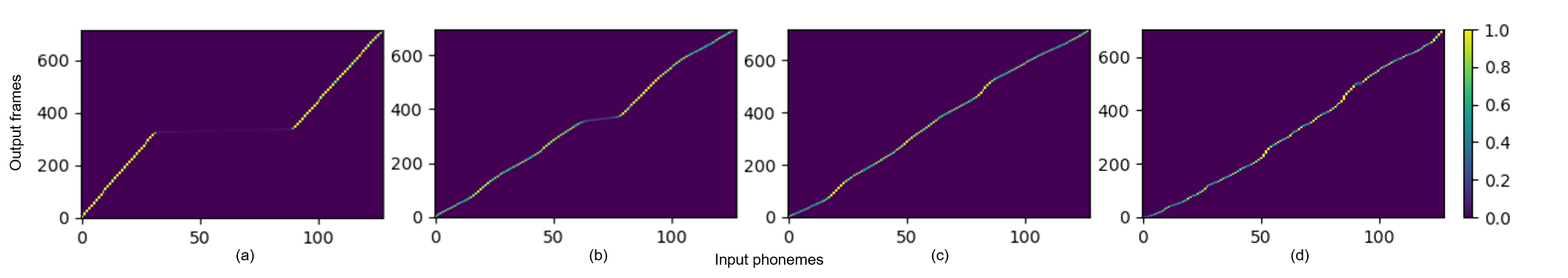}
\end{center}
   \caption{Duration alignments between input phonemes and output frames in our proposed model. The three unsupervised models are compared to the last predicted alignment computed from supervision of target durations. (a) Without Soft-DTW, (b) With Soft-DTW, (c) Phoneme-wise ($\alpha=2$) (d) With duration target
}
\label{fig:3}

\end{figure*}

\section{Reinforce-Aligner}
\subsection{Reinforcement Learning Setup}
The general architecture of the reinforce-aligner is shown in Figure \ref{fig:1}. The duration predictor is the agent that interacts with the environment for each step of the training process. The environment is the overall text-to-waveform network. As shown in the generator training process of Figure \ref{generator}, duration predictions are each upsampled to waveforms, and reward is calculated by mel-spectrogram losses of the respective mel-spectrograms. The reward feedback is given back to the agent for the final action selection.  



\subsection{Agent}
The duration predictor returns one scalar value for each phoneme duration prediction. Predictor consists of two 1D convolutional layers, each with layer normalization, ReLU activation, and dropout following \cite{ren2019fastspeech}. The last linear layer reshapes the convolutional output into a single scalar value. There are two actions available for the agent: 
\begin{itemize}
    \item \textit{KEEP}: Keep phoneme duration prediction without any alterations to the prediction output.
    \item \textit{SHIFT}: Shift phoneme duration prediction with shift value applied in alternating signs. 
\end{itemize}
Shift applied to each phoneme duration consists of alternating signs, which is important to maintain the total sum of the duration prediction outputs. Additionally, we designed two types of shifts: Segments-wise and phoneme-wise shifts. Segment-wise shift corresponds to shift applied to the entire phoneme sequence segments. Phoneme-wise shift is shift applied to each phoneme duration of a phoneme sequence. We examine the effect of shift type and value on the outcome of alignments and speech quality in our ablation study. 

\subsection{Environment}
In this reinforcement learning based setup, the environment is the trained text-to-waveform network that outputs audio waves from phoneme inputs. There are two main objectives of the environment in the reinforce-aligner: (i) Provide input phoneme sequence information to the agent before it takes action, (ii) Give feedback to the agent after considering each of the two possible actions.

\subsubsection{State}
The environment produces phoneme embedding outputs for the agent in each training step. From phoneme sequence inputs, the multi-receptive field fusion based phoneme encoder generates phoneme embedding outputs through multiple residual blocks. These encoder outputs are the current state inputs for the agent to decide on an action.

\subsubsection{Rewards}
We have two different rewards depending on the type of shift. Both rewards consider the mel-spectrogram loss, which is the L1 loss between mel-spectrograms of ground-truth and generated waveforms. As shown in Figure \ref{generator}, each mel-spectrograms are produced by the waveforms synthesized from predicted (\textit{KEEP}) and shifted (\textit{SHIFT}) durations. The segment-wise reward compares loss values of the entire wave segments used for training. Lower loss value implies higher similarity of the generated waveform to the ground-truth. The phoneme-wise reward considers phoneme-wise mel-spectrogram loss values. Specifically, the mel-spectrogram loss values are interpolated by downsampling into the shape of the phoneme duration sequences. Denote the phoneme duration sequences of predicted duration as $D_k = [d_1, d_2, d_3, ..., d_j]$ and shifted duration as $D_s = [d_1+\alpha, d_2-\alpha, d_3+\alpha ..., d_j\pm\alpha]$, where $\alpha$ represents the shift value. Then, our reward is formulated as:                
\begin{equation}
    r_{j}=
    \begin{cases}
        r_{kj}=1 \text{  and  }  r_{sj}=0, & \mbox{if } \mathcal{L}_{kj} \leq \mathcal{L}_{sj} \\
        r_{kj}=0 \text{  and  }  r_{sj}=1, & \mbox{if } \mathcal{L}_{kj} > \mathcal{L}_{sj}
    \end{cases}
\end{equation}
Here, $\mathcal{L}_k$ represents the L1 loss between predicted and ground-truth mel-spectrograms, and $\mathcal{L}_s$ is the L1 loss between shifted and ground-truth mel-spectrograms. $r_k$ and $r_s$ are the keep reward and shift reward values, respectively. Each $j$ is a phoneme duration index value in the phoneme duration sequence for a total of $N$ indices. For the segment-wise reward, keep reward values and shift reward values each have equal values $\forall{j}$. For the phoneme-wise reward, there are unique keep and shift reward values for each $j$. 

\subsection{Gaussian Upsampling}
Predicted durations are scaled to the length of frame sequence outputs. Each scaled predictions are used to find cumulative sum of scaled token lengths and their center positions, as introduced in \cite{donahue2020end}. We first compute the weights:
\begin{equation}
    w_{t}^{i} = \cfrac{\exp({\sigma^{-2}(t-c_{i})^{2})}}
    { \sum_{j=1}^{i}\exp({\sigma^{-2}(t-c_{j})^{2})} }
\end{equation}
given fixed temperature parameter $\sigma^{2}$, scaled
token center position $c$, and time step $t$. We finalize upsampling by producing weighted-sum between the encoder outputs and weights. The output features are used as decoder inputs and transposed convolutions upsample the features to a raw waveform.           

\subsection{Reinforced Duration Loss}
We provide the agent with appropriate feedback by a loss that incorporates the rewards and duration prediction actions. For each phoneme sequence index $j$, the duration values are compared between the original duration prediction and the duration of the selected action. The loss is defined as:
\begin{equation}
    \mathcal{L}_{re} = \sum_{j=1}^{N}(||{D}_{predj}-({D}_{kj} \times {r}_{kj} + {D}_{sj} \times {r}_{sj})||_{1})
\end{equation}
Given $N$ total tokens, each $D_k$, $r_k$ pairs and $D_s$, $r_s$ pairs represent duration and reward values for keep, shift actions, respectively. $D_{pred}$ is equal to $D_k$ because the keep action does not shift the predicted duration values. Therefore, the loss returns positive loss for shift action, and zero loss for keep action.    

\section{Auxiliary Loss}
We utilize GAN loss \cite{mao2017least} and reinforced duration loss. Additionally, auxiliary losses were used to support training of our text-to-wave network.   

\subsection{Total Duration Loss}
The main purpose of the aligner is to produce accurate alignments of phoneme-to-frame sequences. However, the aligner does not have "correct" duration alignments to refer to during training, and therefore is not certain the duration outputs are accurate. Therefore, total duration loss is utilized as guidance for the model. We refer to the aligner length loss in \cite{donahue2020end}. Let $m_{length}$ be the length of ground-truth mel-spectrogram, and $l_{j}$ be the predicted length of $j$th token. The total duration loss is: 
\begin{equation}
    \mathcal{L}_{total} = ({m}_{length}-\sum_{j=1}^{i}{l}_{j})^{2}
\end{equation}

\subsection{Mel-Spectrogram Loss}
In \cite{kong2020hifi}, mel-spectrogram loss is mentioned to be able to optimize the generator and improve quality of generated waveforms. Additionally, reward designs for the reinforce-aligner depends on the mel-spectrogram loss values to produce quality feedback for the agent in our model. The loss is formulated as: 
\begin{equation}
    \mathcal{L}_{mel} = \sum_{t=1}^{T}(||{mel}_{gt}[t]-{mel}_{pred}[t]||_{1})
\end{equation}
where ${mel}_{gt}$, ${mel}_{pred}$ are mel-spectrograms of ground-truth and synthesized waveforms for $T$ time steps.   

\subsection{Soft Dynamic Time Warping}
We enable the mel-spectrograms to have room for error by iteratively finding an alignment path between ground-truth and synthesized spectrograms with dynamic time warping (DTW) \cite{sakoe1971dynamic, donahue2020end}. The main objective of this method is to alleviate the requirement that both spectrograms must be exactly aligned. 

The total cost is defined as:
\begin{equation}
    {c}_{total} = \sum_{l=1}^{L}(\delta_{l} \cdot\omega + \sum_{t=1}^{T}(||{mel}_{gt}[t]-{mel}_{pred}[t]||_{1}))
\end{equation}
where $L$ is the mel-spectrogram length, and $\omega$ is the warp penalty that occurs for actions 2 and 3. $\delta$ is a binary indicator that is 1 when warp penalty is greater than zero. For our loss, we use the soft minimum from \cite{cuturi2017soft} to produce Soft-DTW.  

\begin{table}[!t]
    \caption{MOS with $95\%$ confidence intervals and duration prediction error results of the ablation study.}
    \centering
    \begin{tabular}{l|c|c} \Xhline{3\arrayrulewidth}
       
     \textbf{Setting} &\textbf{MOS} &\textbf{Dur. Error}    \\  \hline
            Ground truth &${4.52}\pm{{0.06}}$ &-   \\ \hline
            w/ duration target &${4.42}\pm{{0.04}}$ &${1.03}$   \\ \hline            
            w/o Soft-DTW &$1.49\pm{0.08}$  &$2.85$  \\ 
            w/ Soft-DTW &$2.91\pm{0.06}$ &$2.53$  \\ 
            Segment-wise ($\alpha=1$)  &$3.48\pm{0.08}$ &$2.24$  \\
            Segment-wise ($\alpha=2$)  &$3.93\pm{0.07}$ &$1.84$  \\
            Phoneme-wise ($\alpha=1$)  &$3.78\pm{0.05}$ &$2.14$  \\
            Phoneme-wise ($\alpha=2$)  &$\textbf{4.06}\pm{\textbf{0.06}}$ &$\textbf{1.78}$  \\
        \Xhline{3\arrayrulewidth}
    \end{tabular}
    
    \label{ablation_table}
\end{table}

\section{Experimental Results and Analysis}
\subsection{Experimental Settings}
For our experiments, we use the National Institute of Korean Language (NIKL) corpus \cite{niklcorpus} and the LJSpeech dataset \cite{ljspeech17}. The NIKL corpus contains about 45K audio samples of 50 native Korean speakers reading Korean text. The LJSpeech dataset contains 13,100 samples recorded by a single English female speaker. For both datasets we unified the sampling rates to 22,050 Hz. In NIKL corpus, we randomly split 10 audio samples for validation, 10 audio samples for testing, and the rest for training in all of the speakers. In LJSpeech dataset, we randomly divided 300 samples for validation, 300 samples for testing, and the rest for training. We conducted our experiments on 4 NVIDIA A100 GPUs.

\subsection{Training Setup}
All experiments are conducted with batch size of 64, using AdamW optimizer \cite{loshchilov2017decoupled} with $\beta_{1}$=0.8, $\beta_{2}$=0.99, and weight decay of $\lambda$=0.01. Learning rate decay of 0.999 was used for each epoch, and the initial learning rate started with 0.0002. We used hidden representation of 256 , frame segment $\gamma$=128, fixed temperature parameter $\sigma^{2}$=10, and other hyper-parameters identical to HiFi-GAN V1, V2 \cite{kong2020hifi} for all models to ensure fairness of experiments. The FFT, window size, and hop size were set to 1024, 1024, and 256. The english texts were converted to phoneme using the method of \cite{g2pE2019}.  All models were trained for 350k steps.


\subsection{Ablation Study}
We conduct ablation study with NIKL corpus to determine the optimal reward for our model. We feed in speaker embedding alongside the encoder output to the reinforce-aligner for training in the multi-speaker dataset. All models are trained with the text-to-waveform network proposed in this paper, and each settings are variations on the alignment method, reward, and shift values. "with duration target" setting represents experiments conducted with attention alignments extracted from Tacotron 2 \cite{shen2018natural}. We synthesize 100 utterances randomly sampled from NIKL corpus test dataset. Afterwards, 12 subjects rated the quality of synthesized speech with scores in the range of 1 (worst) to 5 (best) in increments of 1. MOS score and duration error results are displayed in Table \ref{ablation_table}. "Phoneme-wise ($\alpha=1$)" represents a phoneme-wise reward shifted by scalar value of 1 in alternating signs. In table \ref{ablation_table}, "Phoneme-wise ($\alpha=2$)" showed the highest MOS score and lowest duration error. The duration error is the L1 loss between predicted duration and target duration extracted from attention alignments of Tacotron 2 . Additionally, Figure \ref{fig:3} visualizes the effectiveness of each settings with different alignment methods.

\subsection{Comparison with Other Methods}
We compare our model with other state-of-the-art methods developed for the task of TTS. Both Glow-TTS and BVAE-TTS use internal aligners. In this comparison, we utilize the "Phoneme-wise ($\alpha=2$)" setting, which represents the best MOS from our ablation study. Our model produces raw waveforms directly from text, and other methods require a vocoder to synthesize waveform from produced spectrograms. For the vocoder, HiFi-GAN \cite{kong2020hifi} is used. We conducted a subjective 5 scale MOS test on Amazon Mechanical Turk \cite{mturk}. At least 20 subjects rated naturalness of audio on a scale of 1 to 5 with 1 point increments. In Table \ref{main_table}, we display the MOS score followed by computed MCD$_{13}$ \cite{kubichek1993mel} and RMSE$_{f0}$ \cite{hayashi2017investigation} results. We synthesize 100, 200 utterances for subjective, objective evaluations, respectively. Our model has the highest MOS score, and lowest MCD$_{13}$, RMSE$_{f0}$ values.           

\begin{table}[!t]
    \caption{MOS with $95\%$ confidence intervals, MCD$_{13}$, and RMSE$_{f0}$ results of proposed and state-of-the-art methods.}
    \centering
     \resizebox{1\columnwidth}{!}{
    \begin{tabular}{l|c|c|c} \Xhline{3\arrayrulewidth}
       
     \textbf{Method} &\textbf{MOS} &\textbf{{MCD}$_{13}$}  &\textbf{RMSE$_{f0}$}    \\  \hline
            Ground Truth &${4.32}\pm{{0.03}}$ &-   &-\\ 
            Vocoded \cite{kong2020hifi} &$4.10\pm{0.04}$  &$1.464$  &$38.013$\\ \hline
            Tacotron2 \cite{shen2018natural} &$4.02\pm{0.05}$  &$4.455$  &$39.260$\\
            Glow-TTS \cite{kim2020glow} &$3.93\pm{0.05}$  &$3.944$  &$40.438$\\
            BVAE-TTS \cite{lee2020bidirectional} &$3.70\pm{0.06}$  &$4.002$  &$41.686$\\
            Proposed (Ours) &$\textbf{4.07}\pm{\textbf{0.04}}$  &$\textbf{2.887}$  &$\textbf{38.988}$\\
            
        \Xhline{3\arrayrulewidth}
    \end{tabular}
    }
    \label{main_table}
\end{table}

\section{Conclusion}
 We propose an end-to-end text-to-waveform network with a novel reinforcement learning based duration alignment search method. The advantage of this model is in the agent's ability to actively search for the optimal duration alignment through action based on reward feedback. We conducted a series of experiments to select the optimal reward for our reinforce-aligner. Our proposed model was able to outperform other state-of-the-art methods with more accurate duration alignments and enhanced naturalness of synthesized audio.

\clearpage
\bibliographystyle{IEEEtran}

\bibliography{mybib}


\end{document}